\documentclass[journal]{IEEEtran}

\usepackage{cite}

\ifCLASSINFOpdf
  \usepackage[pdftex]{graphicx}
\else
\fi

\usepackage{amsmath}
\usepackage{amssymb}

\usepackage{siunitx}
\DeclareSIUnit{\pixel}{px}
\renewcommand{\vec}{\boldsymbol}
\DeclareMathOperator*{\argmin}{arg\,min}

\usepackage{adjustbox}
\usepackage{pgf}

\begin{document}
\pagenumbering{gobble}
%
\title{Deep Scatter Splines: Learning-Based Medical X-ray Scatter Estimation Using B-splines}
%
%
%
\author{
    Philipp~Roser, Annette~Birkhold, Alexander~Preuhs, Christopher~Syben, Norbert~Strobel, Markus~Korwarschik, Rebecca~Fahrig, Andreas~Maier%
    \thanks{P.~Roser, A.~Preuhs, C.~Syben, and A.~Maier are with the Pattern Recognition Lab, Department of Computer Science, Friedrich-Alexander Universit\"at Erlangen-N\"urnberg, Erlangen, Germany. P.~Roser is funded by the Erlangen Graduate School in Advanced Optical Technologies (SAOT), Friedrich-Alexander Universit\"at Erlangen-N\"urnberg, Erlangen, Germany. A.~Maier is principal investigator at the SAOT. A.~Birkhold, M.~Kowarschik, and R.~Fahrig are employees of Siemens Healthcare GmbH, 91301 Forchheim, Germany. N.~Strobel is with the Institute of Medical Engineering Schweinfurt, University of Applied Sciences Würzburg‐Schweinfurt, 97421 Schweinfurt, Germany. }%
}%
\maketitle

\begin{abstract}
The idea of replacing hardware by software to compensate for scattered radiation in flat-panel X-ray imaging is well established in the literature.
Recently, deep-learning-based image translation approaches, most notably the U-Net, have emerged for scatter estimation.
These yield considerable improvements over model-based methods. 
Such networks, however, involve potential drawbacks that need to be considered.
First, they are trained in a data-driven fashion without making use of prior knowledge and X-ray physics.
Second, due to their high parameter complexity, the validity of deep neural networks is difficult to assess.
To circumvent these issues, we introduce here a surrogate function to model X-ray scatter distributions that can be expressed by few parameters.
We could show empirically that cubic B-splines are well-suited to model X-ray scatter in the diagnostic energy regime.
Based on these findings, we propose a lean convolutional encoder architecture that extracts local scatter characteristics from X-ray projection images.
These characteristics are embedded into a global context using a constrained weighting matrix yielding spline coefficients that model the scatter distribution.
In a first simulation study with 17 thorax data sets, we could show that our method and the U-Net-based state of the art reach about the same accuracy.
However, we could achieve these comparable outcomes with orders of magnitude fewer parameters while ensuring that not high-frequency information gets manipulated.
\end{abstract}


%
\IEEEpeerreviewmaketitle

\section{Introduction}
\IEEEPARstart{F}{lat-panel} detector cone-beam computed tomography (CBCT) is a powerful technique to enrich fluoroscopically-guided interventions (FGI) with 3-D information of a patient's anatomy.
In contrast to conventional spiral computed tomography (CT), flat-panel X-ray acquisitions suffer, due to the larger X-ray field, from severe photon scattering effects.
Without scatter compensation, reconstructed 3D images are likely to be impaired by streaking or cupping artifacts \cite{Ruehrnschopf:2011:ScatterCompensation1,Ruehrnschopf:2011:ScatterCompensation2}.
In principal, two compensation strategies are possible: hardware-based and software-based approaches.

Hardware-based solutions describe direct manipulations of the X-ray field to either suppress incident scatter or modulate the primary beam.
Typically, today's systems are equipped with anti-scatter grids mounted in front of the detector.
They physically block a large portion of incoming scattered photons \cite{Chan:1985:ASG}.
By increasing the object-detector-distance, scatter can be reduced as well.  
More sophisticated, experimental approaches include primary modulation \cite{Bier:2017:PrimaryModulation} and slit scanning \cite{Bhagtani:2009:Slit}.
In general, and especially for interventional imaging, hardware-based solutions may complicate flexible adjustment of imaging settings or introduce artifacts of their own \cite{Gauntt:2006:ASGLines}.
Therefore, and since hardware-based approaches are often accompanied by supporting algorithms in most cases anyways, purely software-based scatter correction methods are very desirable.

As in many applications, Monte Carlo (MC) simulation is considered the gold standard for scatter estimation \cite{Zbijewski:2004:MCScatter,Poludniowski:2009:MCScatter}.
Its high computational complexity, however, still renders the actual application of MC simulation for scatter estimation in the interventional environment impracticable.
As a solution, the Boltzmann transport equation can be solved directly using finite differences and discretization \cite{Wang:2018:Acuros}.
Yet, an acceptable run-time is only achievable by employing dedicated graphics processing units (GPU).
This is why simpler, model-based approaches are still most widely used in clinical systems.
They aim at estimating scatter from the X-ray projections directly, and are either based on simplified physical or analytical descriptions of X-ray scattering \cite{Yao:2009:AnalyticalModel,Meyer:2009:EllipticModel} or expressed through
convolutional kernels \cite{Ohnesorge:1999:KSE}.
Unfortunately, model-based approaches are likely to fail if acquired images do not meet the underlying assumptions. 

With the advent of convolutional neural networks (CNN) and most notably the U-Net \cite{Ronneberger:2015:Unet} in the context of image-to-image translation, promising advances in the field of X-ray scatter correction were reported \cite{Maier:2019:DSE,Roser:19:DoseLearning}.
The Deep Scatter Estimation (DSE) framework \cite{Maier:2019:DSE} is, for example, able to estimate X-ray scatter solely based on X-ray projections.
In fact, it is on par with MC methods under optimal simulation conditions.
However, the application of deep CNNs for image-to-image translation, especially in diagnostic imaging or FGIs, raises questions that are difficult to answer.
Besides the overall difficulty in comprehending the operating principle of deep networks, the large number of trainable weights makes the U-Net highly dependent on the associated training data and therefore prone to adversarial effects and over-fitting.
In addition, omitting prior knowledge can lead to higher maximum error bounds \cite{Maier:2019:KnownOperators}.
Finally, to maintain a fast run-time, U-Net-based scatter estimation is likely to require a dedicated, high-end GPU potentially adding considerable extra cost to the system.

We believe that by including additional constraints on the scatter characteristics within a CNN-based approach, improvements can be achieved. 
Since in the medical X-ray energy regime the majority of scattered radiation is low-frequency, there needs to exist a low-dimensional scatter representation.
Here, we show that this low-frequency scatter distribution can be approximated by sparse bivariate splines
\cite{Schoenberg:1946:Approximation} with an accuracy of up to \SI{98}{\percent}.
To this end, we train a CNN-based encoder to directly infer this representation from X-ray projections.
Thanks to the spline approximation, the number of parameters to train could be reduced by orders of magnitude, while maintaining the accuracy of the established U-Net.
By using a low-dimensional spline as surrogate, we also ensure low-frequency characteristics of the resulting scatter distribution.
As a consequence, the proposed method cannot alter any high-frequency information because it is simply removing a smoothly varying background.

\section{Material and Methods}
\subsection{X-ray Scatter Model}
In general, three photon interaction mechanisms are relevant for scatter in X-ray projections: (1) Rayleigh scattering, (2) Compton scattering, and (3) arbitrary, subsequent combinations of both, related as multiple scattering.
Although each of them is subject to typical photon Poisson noise, the total scatter distribution in X-ray projections has low-frequency characteristics.
In the following, we consider the flat-field normalized X-ray projection $\vec{I} \in \mathbb{R}^{w\times h}$ with width $w$ and height $h$ in pixels. 
The X-ray projection consists of the primary signal $\vec{I}_\text{p} \in \mathbb{R}^{w\times h}$ described using the Beer-Lambert law and scatter $\vec{I}_\text{s} \in \mathbb{R}^{w\times h}$.
The scatter signal includes all scattering mechanisms.
Thus $\vec{I}$ at pixel $\vec{u} = (u_w, u_h)$ is given by
\begin{equation}
    \vec{I}(\vec{u}) = \underbrace{\int_{E} I_{0}(E) e^{-\int_x \mu(x,E) dx} dE}_{\vec{I}_\text{p}(\vec{u})} + \vec{I}_\text{s}(\vec{u}) \,,
\end{equation}
with the photon energy $E$, X-ray spectrum $I_0(E)$, and the linear attenuation coefficient $\mu(x,E)$.

We approximate $\vec{I}_\text{s}$ at pixel $\vec{u}$ using a bivariate spline of degree $k$ and order $n = k + 1$.
The spline is defined by a $w_c \times h_c$ grid of coefficients $\vec{C} \in \mathbb{R}^{w_c \times h_c}$ on a fixed knot grid $\vec{T} \in \mathbb{R}^{(w_c - k + 1) \times (h_c - k + 1)}$.
This gives us the surrogate $\tilde{\vec{I}}_\text{s,n}(\vec{u})$ at pixel $\vec{u} = (u_0, u_1)$
\begin{equation}
    \tilde{\vec{I}}_\text{s,n}(\vec{u}) = \sum_{i=i_0}^{n+i_0} \sum_{j=j_0}^{n+j_0} \vec{C}_{i,j} B_{i,n,\vec{T}}(u_0) B_{j,n,\vec{T}}(u_1) \,,
\end{equation}
with the basis splines $B$.
Note that the starting indices $i_0$ and $j_0$ depend on both the evaluated pixel $\vec{u}$ and the knot grid $\vec{T}$.
Since $\vec{T}$ is fixed for any X-ray projection, we fit the associated splines in a least-square sense, following Dierckx \cite{Dierckx:1993:CSF}.
The error residual $\varepsilon_r = ||\vec{I}_\text{s} - \tilde{\vec{I}}_\text{s} ||_2$ approaches 0 with increasing number of spline coefficients.

\subsection{Proposed Method}
\label{sec:methods}
\begin{figure}
    \centering
    \includegraphics[width=\linewidth]{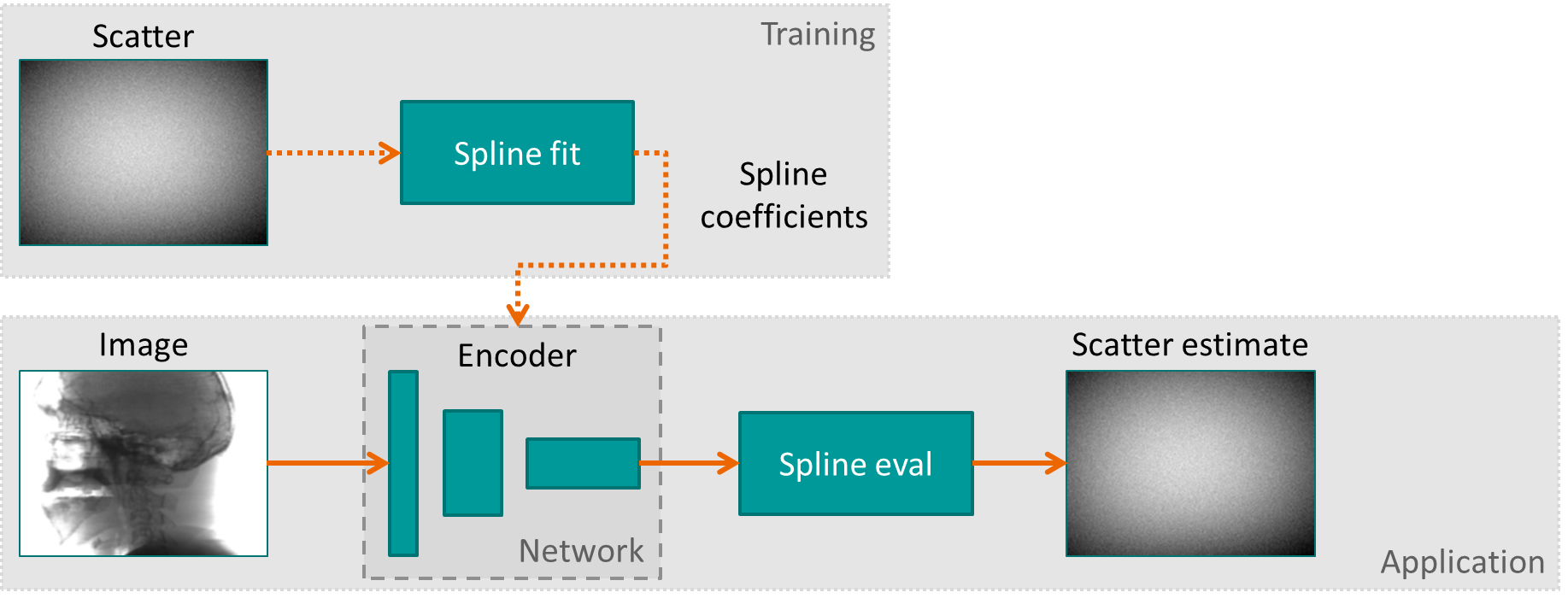}
    \caption{
        Overview of the proposed method.
        Ground truth scatter distributions are approximated using bivariate B-splines.
        Then a convolutional encoder is trained to directly infer spline coefficients from the projection image.
        Afterwards, these spline coefficients are used to calculate the actual scatter distribution of the underlying input image.
    }
    \label{fig:outline}
\end{figure}
Based on our scatter model, we propose to infer scatter spline coefficients directly from the X-ray projection domain.
The outline of the presented method is depicted in Fig.\,\ref{fig:outline}.
From ground truth MC scatter simulations, we find a sparse scatter representation by fitting bivariate splines with respect to an arbitrary but fixed knot grid $\vec{T}$.
The resulting spline coefficients $\vec{C}$ are then used to train a deep encoder to extract such a sparse scatter representations from an X-ray projection.
By evaluating the inferred spline coefficients using the previously defined grid, we are able to obtain the low-frequent portion of scatter distributions for any desired pixel spacing.

The encoder employed consists of blocks of two convolutional layers, each followed by a rectified linear unit (ReLU). 
The number of blocks is arbitrary and manually adjustable to the input and output dimensions.
Between two convolutional blocks, the dimensions of the feature maps are halved by averaging a neighborhood of \num{2x2} pixels for each feature map.
The number of intermediate feature maps is constant for all levels.
Before each convolution, replication padding is applied to preserve spatial dimensions.
The last block is followed by a weighted sum over all channels to reduce the number of intermediate channels to one.
Eventually, the latent variable of width $w_l$ and height $h_l$ is vectorized.
This convolutional path $\mathcal{S}_{\vec{\theta}}: \mathbb{R}^{w \times h} \mapsto \mathbb{R}^{w_l \cdot h_l \times 1}$ extracts latent variables that locally characterize X-ray scattering.
However, scattered radiation globally affects image formation.
To account for this, we model each spline coefficient $C_{i,j}$ as the weighted, normalized sum of all intermediate latent variables.
This global relationship can be expressed by multiplying the latent variable with matrix $\vec{W} \in \mathbb{R}^{w_c \cdot h_c \times w_l \cdot h_l}$, which is only feasible due to the low dimensionality of the latent space and not directly transferable to the DSE.

To be comparable to the DSE, the network parameters $\vec{\theta}$ and $\vec{W}$ are optimized by minimizing the mean absolute percentage error (MAPE) with respect to the ground truth
\begin{equation}
   \vec{\theta}, \vec{W} = \argmin_{\vec{\theta}, \vec{W}}  \sum_i^{w_c} \sum_j^{h_c} \frac{||\left[\vec{W}\mathcal{S}_{\vec{\theta}}(\vec{I})\right]_{i,j} - \vec{C}_{i,j}||_1}{\vec{C}_{i,j}},
\end{equation}
where $\left[\cdot\right]_{i,j}$ is the access operation for a vectorized matrix.
Since $\vec{W}$ models a weighted sum, we impose two constraints: (1) each element of $\vec{W}$ must be positive $W_{i,j} > 0, \forall i, j$ and (2) each row sums up to unity $\sum_j W_{i,j} = 1, \forall i$.

\subsection{Ground Truth Data Simulation}
Acquiring matching pairs of training data using conventional X-ray imaging is tedious and time-consuming.
As an alternative, we used the X-ray transport code MC-GPU \cite{Badal:09:MCGPU} to generate synthetic training data.
As inputs for our simulation, openly accessible CT scans from The Cancer Imaging Archive (TCIA) \cite{Clark:13:TCIA} were used.
In total, we extracted 17 thorax phantoms from the CT Lymph Nodes torso \cite{Roth:15:Lymphs} data sets.
To prepare the phantoms for MC simulation, we employed a basic pre-processing pipeline \cite{Roser:20:MCPreProcessing} to remove non-patient objects, and assign tissues and densities to each voxel. 
Based on the simulated scatter distributions, we fit bivariate cubic B-splines.

For each X-ray projection, we simulated \num{5e10} primary photons sampled from a \SI{100}{\kilo\eV} tungsten spectrum.
In total, we considered three slightly different iso-center positions $\vec{o} \in \mathbb{R}^3$, seven different projection angles $\alpha \in \{0, 15, 30, \dots, 90\}$ in \si{\degree}, and fixed source-to-detector (\SI{1300}{\mm}) and source-to-iso-center (\SI{800}{\mm}) distances, respectively.
The virtual detector is \SI{399 x 299}{\mm} large and consists of \num{1248 x 928} pixels.
To reduce simulation noise, we filter the scatter signals using a Gaussian kernel $\vec{G}_\sigma$.

\section{Results}
\label{sec:results}
\subsection{Spline Approximation Capability}
\begin{figure}
    \centering
    \resizebox{0.95\linewidth}{!}{
        \begin{adjustbox}{trim=20pt 0 0 0}
        \input{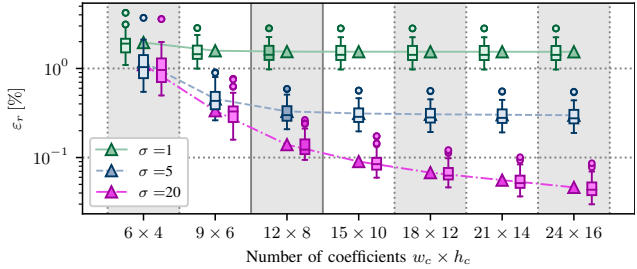}
        \end{adjustbox}
    }
    \caption{
        Absolute average spline approximation error $\varepsilon_r$ (triangles) and associated box plots for different numbers of coefficients for strong ($\sigma = 20$), medium ($\sigma = 5$), and no ($\sigma = 1$) Gaussian smoothing.
    }
    \label{fig:results:spline-capability}
\end{figure}

Initially, we investigated the quality of splines for medical X-ray scatter approximation.
To this end, we fit cubic splines ($k = 3$) with a different number of coefficients $\vec{C} \in \mathbb{R}^{w_c \times h_c}$ to the scatter signal and juxtaposed the associated residuals $\varepsilon_r$.
Figure \ref{fig:results:spline-capability} summarizes the results for $w_c \times h_c \in \{48 \times 32, 24 \times 16, 12 \times 8, 6 \times 4, 3 \times 2\}$ for all considered simulated X-ray scatter signals.
Evidently, only marginal improvements can be observed with increasing number of coefficients.
The best trade-off between sparsity and a low approximation error was obtained with $12 \times 8$ coefficients.
Hence, we considered $\vec{C} \in \mathbb{R}^{12 \times 8}$ for further analysis.

\subsection{Experimental Setup}
\begin{table}
    \centering
    \caption{Number of parameters to train and quantitative test results for all investigated network architectures for different depths $d$ and number of feature maps $c$. For better assessment, the median absolute percentage error (MedAPE) is also given.}
    \label{tab:results}
    \bgroup
    \def\arraystretch{1.2}%
    \begin{tabular}{lccccc}
    \hline\hline
        Architecture & d & c & Parameters & MAPE [\si{\percent}] & MedAPE [\si{\percent}]\\
        \hline
        U-Net
        & 6
        & \num{8} & \num{12.6}\,M & \num{4.66\pm 4.09} & \num{3.59} \\
         & & \num{16} & \num{50.3}\,M & \num{5.06\pm 3.89}  & \num{4.24}\\
        & 5
        & \num{8} & \num{3.1}\,M & \num{5.04\pm 4.25} & \num{3.97} \\
        & & \num{16} & \num{12.6}\,M & \num{5.41\pm 4.13}  & \num{4.60}\\
        
        \hline
        
        Ours
        & 7
        & \num{8} & \num{10.1}\,k & \num{4.84\pm 4.01} & \num{3.80} \\
        & & \num{16} & \num{32.7}\,k & \num{4.53\pm 3.82} & \num{3.51} \\
        & 6
        & \num{8} & \num{15.8}\,k & \num{4.66\pm 4.22} & \num{3.55} \\
        & & \num{16} & \num{35.0}\,k & \num{4.67\pm 4.29} & \num{3.65} \\
    \hline\hline
    \end{tabular}
    \egroup
\end{table}
In accordance to the DSE \cite{Maier:2019:DSE}, our encoder network is trained with down-sampled X-ray projections comprising \num{384 x 256} pixels and targets \num{12 x 8} spline coefficients.
We excluded one patient for validation during training (\num{1x21} X-ray projections) and one patient for testing afterwards (\num{2x21} X-ray projections).
Hence, the encoder was trained using \num{273} X-ray projections from \num{13} patients and a batch size of \num{16}.
For data augmentation, we used random horizontal flips and random Poisson noise as these techniques preserve physical plausibility.
The network was trained using adaptive moments optimizer with default parameters \cite{Kingma:2014:Adam} subject to minimize the MAPE to the target MC scatter simulations.
The results were compared to the U-Net with respect to the ground truth MC scatter simulations.
We trained the baseline U-Net and our network architecture with different depths $d$ and number of initial feature maps $c$ (see \,Tab.\,\ref{tab:results}).

\subsection{Comparison of the Network Architectures}
Table\,\ref{tab:results} shows the pixel-wise MAPE with respective standard deviations and the median absolute percentage error (MedAPE) of the investigated network architectures on the two test patients.
Evidently, both network architectures perform similarly for all parameter combinations and median error rates range from \SIrange{3.55}{4.60}{\percent}. 
However, slight trends can be observed.
The U-Net configurations with fewer parameters (8 initial feature maps) yield \SI{0.64}{\percent} lower median error rates on average compared to their high-parameters counterparts.
Such a trend is not seen for our network architecture.
Furthermore, it reaches overall lower error rates on average compared to the U-Net.
Most strikingly, our method is capable to maintain the U-Net's accuracy while reducing the number of parameters to train by two to three orders of magnitude.

\begin{figure}
    \centering
    \setlength\tabcolsep{1.5pt}
    \begin{tabular}{ccccc}
        & Input & Ground Truth & U-Net APE & Ours APE \\ 
         \rotatebox{90}{\small Low} & 
         \includegraphics[width=0.23\linewidth]{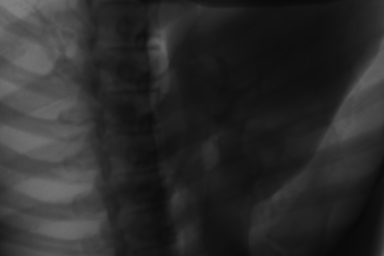} &
         \includegraphics[width=0.23\linewidth]{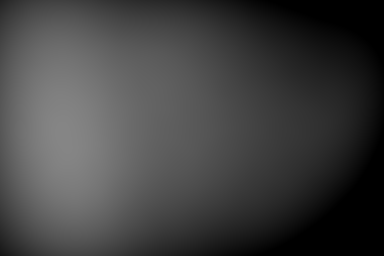} &
         \includegraphics[width=0.23\linewidth]{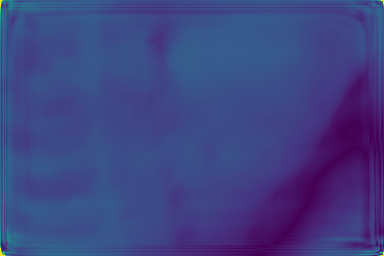} &
         \includegraphics[width=0.23\linewidth]{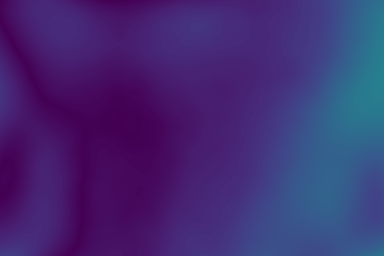} \\
         \rotatebox{90}{\small Normal} & 
         \includegraphics[width=0.23\linewidth]{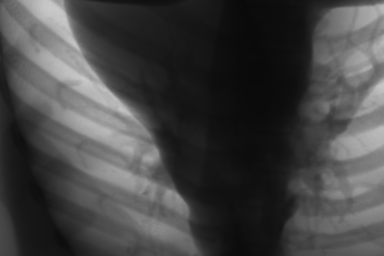} &
         \includegraphics[width=0.23\linewidth]{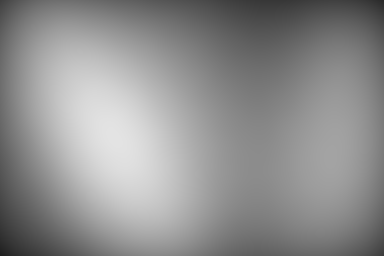} &
         \includegraphics[width=0.23\linewidth]{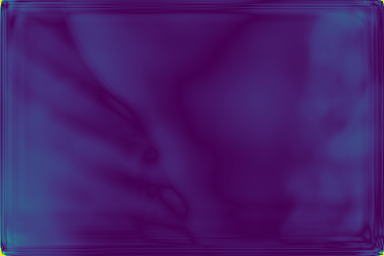} &
         \includegraphics[width=0.23\linewidth]{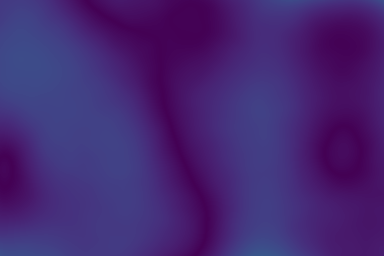} \\
         \rotatebox{90}{\small High} & 
         \includegraphics[width=0.23\linewidth]{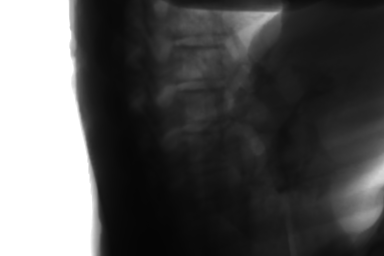} &
         \includegraphics[width=0.23\linewidth]{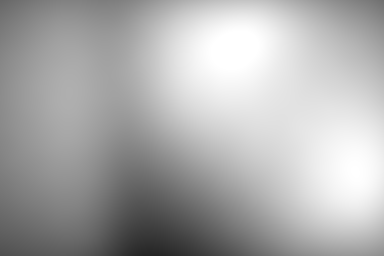} &
         \includegraphics[width=0.23\linewidth]{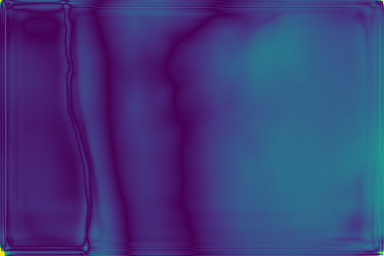} &
         \includegraphics[width=0.23\linewidth]{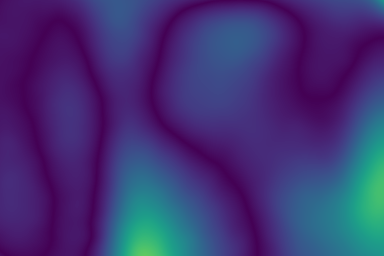} \\
    \end{tabular}
    \caption{Qualitative comparison between both approaches for three different views of different intensity ranges (low, normal, high). Input and ground truth are windowed to give a good visual representation. Absolute percentage errors (APE) are in the range of \SIrange{0}{25}{\percent} (dark blue to yellow).}
    \label{fig:results}
\end{figure}
To assess the network architectures from a qualitative point of view, Fig.\,\ref{fig:results} depicts three example projections, their associated scatter signals, and the absolute percentage errors (APE) of the predictions by the U-Net ($d=6, c=8$) and our encoder ($d=7, c=16$) with respect to the ground truth.
Overall, the scatter distributions are smooth and almost no salient features are observable, in particular in the proposed approach.

\section{Discussion and Conclusion}
\label{sec:discussion}
We developed a novel method to estimate X-ray scatter from projection images in a data-driven fashion.
Leveraging the ability of B-splines to approximate low-frequency signals, we were able to tremendously reduce the number of training parameters to arrive at results comparable to the current state of the art.
We presented a proof of concept based on a simulation study involving 17 thorax data sets at one fixed peak tube voltage relevant for medical X-ray diagnostics.

Although these preliminary results appear promising, several limitations of our study need to be discussed.
First, as advanced pre-processing has shown potential to improve learning-based scatter estimation by approximately \SI{4}{\percent} \cite{Maier:2019:DSE} as well, future studies need to incorporate different pre-processing techniques in the evaluation.
Second, at this point in time we cannot yet make a statement on how 3D reconstruction benefits, as associated experimental results are not yet available.
Third, in contrast to the DSE \cite{Maier:2019:DSE}, we only considered one anatomic region and one tube peak voltage.
Therefore, no assumption on our method's generalizability to other anatomic regions and X-ray energy regimes can be made.
It cannot be ruled out that our lean encoder architecture does not have the same capacity as the U-Net.

Besides these potential limitations, our method comes with interesting characteristics that are very desirable in the interventional environment.
First, the parameter reduction leads to a lower computational complexity.
This allows for (a) accelerated reconstruction, (b) seamless transition between 3D and 2D imaging, and (c) it can be implemented on a wide variety of processing hardware.
In addition, it is questionable whether a single network is needed that can adapt to highly varying tube voltages and anatomic regions.
Multiple dedicated networks could be used instead.
Second, by using a sparse spline as surrogate signal, our method ensures that only low frequencies get altered in the X-ray projection image.
Hence, it is very unlikely that potentially diagnostically relevant anatomic features get removed.

In sum, the use of splines as X-ray scatter surrogate signal has the potential to improve learning-based X-ray scatter estimation algorithms, as it requires significantly fewer parameters and much reduced computational complexity.

~\\
\textbf{Disclaimer:} The concepts and information presented are based on research and are not commercially available.

\ifCLASSOPTIONcaptionsoff
  \newpage
\fi



\bibliographystyle{IEEEtran}
\bibliography{references}
\end{document}